\documentclass{article}

\usepackage{lastpage}
\usepackage{epsfig}

\makeatletter
\def\@oddfoot{\hfill}
\newcount\shumeicount
\def\setshumei#1#2#3{%
  \shumeicount=\count0
  \def\@oddhead{%
    \raise-5pt\hbox to0pt{\vrule width\hsize height 0pt depth 0.4pt\hss}\relax
    \ifnum \shumeicount=\count0
      \raise-7pt\hbox to0pt{\vrule width\hsize height 0pt depth 0.4pt\hss}\relax
      #1
    \else
      \ifodd\count0
        #2
      \else
        #3
       \fi
     \fi
  }%
}
\makeatother

\makeatletter
\def\@oddfoot{\hfill}
\newcount\shujiaocount
\def\setshujiao{%
  \shujiaocount=\count0
  \def\@oddfoot{%
      \ifodd\count0
      \else
      \fi
  }%
}
\makeatother

\def\biaoti#1#2#3#4{{
  \vspace*{0.3cm}
  \begin{flushleft} \Large\bf #1\end{flushleft}
  \vspace*{-0.2cm}
      \begin{flushleft}
      \bf #2
      \end{flushleft}
      \footnotetext{\hspace{-6mm} #3\\ #4}}}
\def\dshm#1#2#3#4
{\setshumei{ {} {}\hfill}
            {\hfill {\small #3}\hfill\hbox to0pt{\hss\thepage}}
            {\hbox to0pt{\thepage\hss}\hfill {\small #4}\hfill
            }
            \setshujiao}
\def\drd#1
{{\vskip 1cm\small \begin{flushleft}
 #1 \\
\copyright 2006 Springer Science + Business Media, Inc.
\end{flushleft}}}
\def\dab#1#2{\noindent {{\small\bf Abstract~~}}{{\small #1}}
            \vskip 0.1cm
             \noindent{{\small\bf Key words~~}}{{\small #2}}
                      }
\def\dse#1{\vskip 0.6cm\noindent
        {\large\bf #1}
        \vskip 0.4cm}
\def\rfne{\vskip 0.5cm  \centerline{\bf References} \vskip 0.5cm
               \parindent 0pt}

\makeatletter
    
    \newcommand{\Rmnum}[1]{\expandafter\@slowromancap\romannumeral #1@}
  \makeatother

\usepackage {amssymb}
\usepackage {amsmath}
\def\dse#1{\vskip 0.6cm\noindent
        {\large\bf #1}
        \vskip 0.4cm}
\begin{document}

%*************************************************************************************************************
\biaoti{New quantum codes from   dual-containing cyclic codes over
finite rings}{Yongsheng Tang$^{1}$, Shixin Zhu$^{2}$,  Xiaoshan
Kai$^{2}$,  Jian Ding$^{3}$ } {1.{\it School of Mathematics and
Statistics, Hefei Normal
University, Hefei, Anhui} $230601$, {\it China}.\\
2.{\it School of Mathematics, Hefei University of Technology, Hefei,
Anhui} $230009$,{\it China}. \\3.{\it Department of Common Course,
Anhui Xinhua University, Hefei, Anhui} $230088$,{\it China}.
\\Email: ysh$_{-}$tang@163.com(Y.Tang),sxinzhu@tom.com(S.Zhu),kxs$6$@sina.com(X.Kai),415134412@qq.com(J.Ding).} {$^*$This research is supported by National Natural Science Funds of
China (Nos. 61370089 and 61572168), Natural Science Foundation of
Anhui Province (No. 1408085QF116), National Mobil Communications
Research Laboratory, Southeast University( No. 2014D04), Colleges
Outstanding Young Talents Program in 2014, Anhui Province ( No.
[2014]181), Anhui Province Natural Science Research (No. KJ2015A308)
and Hefei Normal University Research Project (No. 2015JG09).}

%*************************************************************************************************************
%The submission date of your article. For example: \drd{Received: June 8, 2006}
%*************************************************************************************************************

%*************************************************************************************************************
% The page header of the article.
% \dshm{Year}{Volume}{The capitalized RUNNING HEAD of your article with less than 48 letters}{The capitalized
% AUTHORS list with $\cdot$ separating different names or one can type "The name of the first author et al."
% if there are more than 4 authors.}
%*************************************************************************************************************

\dshm{}{}{}{}

%*************************************************************************************************************
\dab{ Let
$R=\mathbb{F}_{2^{m}}+u\mathbb{F}_{2^{m}}+\cdots+u^{k}\mathbb{F}_{2^{m}}$
, where
 $\mathbb{F}_{2^{m}}$ is the finite
 field with  $2^{m}$ elements,  $m$ is a positive integer, and $u$ is an indeterminate with
 $u^{k+1}=0.$  In this paper, we propose the
constructions of two new families of  quantum codes obtained from
dual-containing cyclic codes of odd length over $R$. A new Gray map
over $R$ is defined and a sufficient and necessary condition for the
existence
 of dual-containing cyclic codes over $R$ is given. A new family of  $2^{m}$-ary quantum codes is
 obtained via the Gray map and the Calderbank-Shor-Steane construction from dual-containing cyclic codes over $R.$
In particular, a new family of binary quantum codes is obtained via
the Gray map, the trace map and the Calderbank-Shor-Steane
construction from dual-containing cyclic codes over $R.$ } {Quantum
codes; dual-containing cyclic
codes; Gray map; trace map}\\

%---------------------------------------------------------------------------------------------------------------
%*************************************************************************************************************

\dse{1~~Introduction}

\label{intro}Quantum codes play an important role not only in
quantum communication but also in quantum computation. Since quantum
codes provide a guarantee for quantum computation and quantum
communication, the construction of quantum codes with good
parameters has been an important topic in quantum information. After
the work of Calderbank et al.$^{[1]}$ gave a thorough discussion of
the principles of quantum codes, the most of the constructions of
quantum codes  were transformed to construct self-orthogonal (or
dual-containing) classical codes. Many works have been done for
constructing good binary quantum codes by using classical codes,
such as Bose-Chaudhuri-Hocquenghem (BCH) codes, Reed-Solomon codes,
Reed-Muller codes and algebraic geometric codes (see Refs. 2-6).
Some of these constructions were generalized to the case of
nonbinary quantum codes, since  nonbinary quantum codes can be
applied in fault-tolerant quantum computation. Ashikhmin and
Knill$^{[7]}$ constructed nonbinary quantum codes by using classical
self-orthogonal codes. Ketkar et al.$^{[8]}$ constructed many
families of nonbinary quantum codes, including quantum Hamming
codes, quadratic residue codes, quantum Melas codes, quantum BCH
codes, and quantum character codes. La Guardia$^{[9,10]}$
constructed new families of nonbinary quantum codes derived from
classical BCH codes.\\

Cyclic codes are a well-studied class of codes that have play an
essential role in both theory and practice. Cyclic codes have also
extensively been used to construct quantum codes. Thangaraj and
McLaughlin$^{[11]}$ constructed   quantum codes from cyclic codes
over $\textrm{GF}(4^{m}).$  R. Li and X. Li$^{[12]}$ constructed
binary quantum codes from cyclic code of length $2^\alpha n$ with
$n$ odd and $n\leq 99$, and $\alpha\leq 2.$  Qian et al.$^{[13]}$
constructed quantum codes from quasi-cyclic  codes. Kai et
al.$^{[14-16]}$ constructed many classes of  quantum codes from
negacyclic and constacyclic codes. Wang and Zhu$^{[17]}$ constructed
non-binary quantum from repeated-root cyclic codes. Chen et
al.$^{[18]}$ constructed quantum codes
from  constacyclic codes.\\

Recently, many good quantum codes have been constructed from cyclic
codes over finite rings.  Qian and Zhang$^{[19]}$ constructed
quantum codes from cyclic codes over
$\mathbb{F}_{2}+u\mathbb{F}_{2}.$ Kai and Zhu$^{[20]}$ constructed
quantum codes from cyclic codes over
$\mathbb{F}_{4}+u\mathbb{F}_{4}.$ Guenda and Gulliver$^{[21]}$
constructed quantum codes from linear codes over finite commutative
Frobenius rings.  Ashraf and Mohammad$^{[22]}$ constructed quantum
codes from cyclic codes over $\mathbb{F}_{3}+v\mathbb{F}_{3}.$ \\

The purpose of this paper is to construct two new families of
quantum codes  by taking advantage of dual-containing cyclic codes
over the finite commutative ring
$\mathbb{F}_{2^{m}}+u\mathbb{F}_{2^{m}}+\cdots+u^{k}\mathbb{F}_{2^{m}}.$
This paper is organized as follows. In Section 2, some definitions
and notations about linear and cyclic codes over
$\mathbb{F}_{2^{m}}+u\mathbb{F}_{2^{m}}+\cdots+u^{k}\mathbb{F}_{2^{m}}$
are provided. In Section 3,  a new Gray map on
$\mathbb{F}_{2^{m}}+u\mathbb{F}_{2^{m}}+\cdots+u^{k}\mathbb{F}_{2^{m}}$
is defined. In Section 4, a necessary and sufficient condition for
the existence of dual-containing cyclic codes is obtained and  a new
family of $2^{m}$-ary quantum codes is constructed.  In Section 5, a
new family of binary quantum codes is constructed. Section 6
concludes the paper.\\
{\hspace*{\parindent}}

\dse{2~~Preliminaries}

\label{sec:1}  Let $\mathbb{F}_{2^{m}}$ be the finite field with
$2^{m}$ elements and $R$ be the commutative ring
$\mathbb{F}_{2^{m}}+u\mathbb{F}_{2^{m}}+\cdots
+u^{k}\mathbb{F}_{2^{m}}\simeq \mathbb{F}_{2^{m}}[u]/(u^{k+1}).$ The
ring $R$   is a finite chain ring whose ideals can be linearly
ordered by inclusion; that is, $0=u^{k+1}R \subset u^{k}R
\subset\cdots \subset u^{2}R \subset uR \subset R.$ The ring $R$ is
given by the obvious addition and multiplication with $u^{k+1}=0$.
The units of $R$ are the elements $a\in R$ such that $a \not\equiv 0
(\textrm{mod} \ u)$ and the residue field is $R/uR \simeq
\mathbb{F}_{2^{m}}.$ A code of length $n$ over $R$ is a nonempty
subset of $R^n$, and a code is linear over $R$ if it is an
$R$-submodule of $R^n$. Given two vectors $\textbf{x}=(x_0,
x_1,\ldots,x_{n-1})$, and $\textbf{y}=(y_0, y_1,\ldots,y_{n-1})\in
R^n$,
 their Euclidean inner product is defined as $$ \textbf{x}\cdot \textbf{y}=x_0y_0+x_1y_1+\cdots+x_{n-1}y_{n-1}\in R.$$
The vectors $\textbf{x}$ and $\textbf{y}$ are called orthogonal with
respect to the Euclidean inner product if $\textbf{x}\cdot
\textbf{y}=0.$ For a  linear code $\mathcal{C}$ of length $n$, the
Euclidean dual code of $\mathcal{C}$ is defined as
$$\mathcal{C}^{\bot}=\{\textbf{x}\in R^n|\textbf{x}\cdot \textbf{y}=0\ for\ all\  \textbf{y}\in \mathcal{C}\}.$$ A
linear code $\mathcal{C}$ of length $n$ over $R$ is called
dual-containing if $\mathcal{C}^\bot \subseteq \mathcal{C}$, and it
is called self-dual if $\mathcal{C}=\mathcal{C}^\bot.$
 Two codes are equivalent if one
can be obtained from the other by permuting the coordinates. Any
code over $R$ is permutation equivalent to a code $\mathcal{C}$ with
generator matrix of the form
$$G=
\left
(\begin {array} {ccccccc} I_{l_{0}} & A_{1,1} & A_{1,2} & A_{1,3} & \cdots & A_{1,k-1} & A_{1,k}\\
0 & uI_{l_{1}} & uA_{2,1} & uA_{2,2}  & \cdots & uA_{2,k-1} & uA_{2,k}\\
0 & 0 & u^{2}I_{l_{2}} & u^{2}A_{3,1}  & \cdots & u^{2}A_{3,k-1} & u^{2}A_{3,k}\\
\cdot & \cdot & \cdot & \cdot & \cdots & \cdot & \cdot\\
0 & 0 & 0 & 0 & \cdots & u^{k}I_{l_{k}} & u^{k}A_{k,1} \\
\end {array}\right),$$
where  $l_{i}$ are all positive integers, $A_{i,j}$ are all matrices
over $R$. Then $\mathcal{C}$ is an abelian group of type
$\{l_{0},l_{1},\ldots,l_{k}\}$, $\mathcal{C}$ contains
$(2^{(k+1)m})^{l_{0}}(2^{km})^{l_{1}}\cdots(2^{m})^{l_{k}}$
codewords, and $\mathcal{C}$ is a free $R$-module if and only if
$l_{1}=l_{2}=\cdots=l_{k}=0$. A linear code $\mathcal{C}$ of length
$n$ over $R$ is called cyclic if it is invariant under the cyclic
shift $\tau$ of $R^n$:
$$\tau(c_0,c_1,\ldots,c_{n-1})=(c_{n-1},c_0,\ldots,c_{n-2}).$$
Each codeword $\textbf{c}=(c_0,c_1,\ldots,c_{n-1})\in \mathcal{C}$
is customarily identified with its polynomial representation
$c(x)=c_0+c_1x+\cdots+c_{n-1}x^{n-1}$, and the code $\mathcal{C}$ is
in turn identified with the set of all polynomial representations of
its codewords. Then in the ring $R[x]/{\langle x^n-1\rangle},$
$xc(x)$ corresponds to a cyclic shift of $c(x)$. It is well known
that a linear code $\mathcal{C}$ of length $n$ over $R$ is cyclic if
and only if $\mathcal{C}$ is an ideal of the quotient ring $R[x]/
{\langle x^n-1\rangle}.$ Throughout this paper, we assume that the
length $n$ is odd. It has been shown that the ring $R[x]/\langle
x^{n}-1\rangle$ is a
principal ideal ring.\\

\dse{3~~Gray Map on $R$  }

Every element $c\in R$ can be written uniquely as
$$c=\beta_{0}(c) +u \beta_{1}(c)+\cdots+u^{k}\beta_{k}(c),$$ where $\beta_{i}(c)\in \mathbb{F}_{2^{m}},$ for
$i=0,1,\ldots,k.$ A new Gray map $\Phi$ on $R$ is defined as
$$\Phi : R  \rightarrow  \mathbb{F}_{2^{m}}^{k+1},$$
$$\beta_{0}(c) +u \beta_{1}(c)+\cdots+u^{k}\beta_{k}(c) \mapsto
(\beta_{k}(c),\beta_{k}(c)+\beta_{0}(c),
\beta_{k-1}(c)+\beta_{0}(c),\beta_{k-1}(c)+\beta_{1}(c),
\cdots,e(c)),$$ where
\begin{eqnarray*}
e(c)=\left\{{{\begin{array}{ll}
 {\beta_{t+1}(c)+\beta_{t}(c),} & {if\  k = 2t+1,} \\
 {\beta_{t}(c)+\beta_{t-1}(c),} & {if\  k = 2t.}\\
\end{array} }}\right.
\end{eqnarray*}
The Gray weight $w_{G}$ of $ c \in R$ is defined  to be the sum of
the Hamming weight of $\Phi (c),$ i.e., $w_{G}(c)=w_{H}(\Phi (c)).$
This extends to a weight function in $ R^{n}:$ if
$\textbf{c}=(c_{0},c_{1},\ldots, c_{n-1}),$ then
$w_{G}(\textbf{c})=\sum_{i=0}^{n-1}w_{G}(c_{i}).$ The Gray distance
$d_{G}(\textbf{c}, \textbf{c}')$ between any distinct vectors
$\textbf{c}, \textbf{c}' \in R^{n}$ is defined to be
$w_{G}(\textbf{c}-\textbf{c}').$ The minimum Gray distance of
$\mathcal{C}$ is the smallest nonzero Gray distance between all
pairs of distinct codewords of $\mathcal{C}$. The minimum Gray
weight of $\mathcal{C}$ is the smallest nonzero Gray weight among
all codewords of $\mathcal{C}$. If $\mathcal{C}$ is linear, then the
minimum Gray distance is the same as the minimum Gray weight. The
Hamming weight $w_{H} $ of a codeword $\textbf{c}\in \mathcal{C}$ is
the number of its nonzero components. The Hamming distance
$d_{H}(\textbf{c}, \textbf{c}')$ between two codewords $\textbf{c}$
and $\textbf{c}'$ is the Hamming weight of the codeword
$\textbf{c}-\textbf{c}'$. The minimum Hamming distance $d_H$ of
$\mathcal{C}$ is defined as $\min\{d(\textbf{c},
\textbf{c}')|\textbf{c},\textbf{c}'\in \mathcal{C}, \textbf{c}\neq
\textbf{c}'\}$. For any linear code, the minimum Hamming distance
$d_H(\mathcal{C})$ of $\mathcal{C}$ is its minimum Hamming weight.
 It is clear that
$\Phi $ preserves linearity.   The Gray map $\Phi$ can be extended
to $R^{n}$ in an obvious way and the extended $\Phi$ is a bijection
from $R^{n}$ to $\mathbb{F}_{2^{m}}^{(k+1)n}.$
\\

The following property of
the Gray map is obvious from the above definitions.\\
\noindent{\bf Proposition 3.1}~~\emph{The map $\Phi$ is a weight
preserving map from $( R^{n},$ Gray weight) to
$(\mathbb{F}_{2^{m}}^{(k+1)n},$ Hamming weight) and a distance
preserving map from $( R^{n},$ Gray
distance) to $(\mathbb{F}_{2^{m}}^{(k+1)n},$ Hamming distance).}\\

 \noindent{\bf Proposition 3.2}~~ \emph{Let $\mathcal{C}$ be a linear code of
length $n$  and type $\{l_{0},l_{1},\ldots,l_{k}\}$ over $R$, and
let $\mathcal{C}^{\perp}$ be the dual of the code $\mathcal{C}$.
Then}
$$\Phi(\mathcal{C})^{\perp}=\Phi(\mathcal{C}^{\perp}).$$

\noindent {\it \textbf{Proof}}~~   Let $ c_{1}=\beta_{0}(c_{1}) +u
\beta_{1}(c_{1})+\cdots+u^{k}\beta_{k}(c_{1})\in \mathcal{C},\
c_{2}=\beta_{0}(c_{2}) +u
\beta_{1}(c_{2})+\cdots+u^{k}\beta_{k}(c_{2})\in
\mathcal{C}^{\perp}.$ Then $c_{1}\cdot c_{2}=0$; that is,
$\beta_{0}(c_{1})\beta_{0}(c_{2})=
\beta_{1}(c_{1})\beta_{0}(c_{2})+\beta_{0}(c_{1})\beta_{1}(c_{2})=\cdots=\sum_{i=0}^{t}\beta_{i}(c_{1})\beta_{t-i}(c_{2})=0,$
for $t=0,1,\ldots,k.$ Therefore
 \begin{eqnarray*}
 \Phi(c_{1})\cdot \Phi(c_{2})&=&(\beta_{k}(c_{1}),\beta_{k}(c_{1})+\beta_{0}(c_{1}),
\beta_{k-1}(c_{1})+\beta_{0}(c_{1}),\beta_{k-1}(c_{1})+\beta_{1}(c_{1}),
\cdots,e(c_{1}))
 \\&&\cdot
 (\beta_{k}(c_{2}),\beta_{k}(c_{2})+\beta_{0}(c_{2}),
\beta_{k-1}(c_{2})+\beta_{0}(c_{2}),\beta_{k-1}(c_{2})+\beta_{1}(c_{2}),
\cdots,e(c_{2}))
 \\
  &=&\sum_{i=0}^{k}\beta_{i}(c_{1})\beta_{k-i}(c_{2})+\sum_{i=0}^{k-1}\beta_{i}(c_{1})\beta_{k-1-i}(c_{2})\\
 &=&0.
 \end{eqnarray*}
Since $\Phi$ is a bijection, it follows that
$\Phi(\mathcal{C}^{\perp})\subseteq \Phi(\mathcal{C})^{\perp}.$
 Now it is enough to show that the two sets have the same
cardinality. Suppose $\mathcal{C}$ is a linear code of length $n$
and type $\{l_{0},l_{1},\ldots,l_{k}\}$ over $R$, we can get
$\Phi(\mathcal{C})$ is a
$[(k+1)n,(k+1)l_{0}+kl_{1}+\cdots+l_{k}]$-linear code over
$\mathbb{F}_{2^{m}}.$ This implies that $\Phi(\mathcal{C})^{\perp}$
is a $[ (k+1)n,(k+1)n-((k+1)l_{0}+kl_{1}+\cdots+l_{k})]$-linear code
over $\mathbb{F}_{2^{m}}.$ Therefore,
$|\Phi(\mathcal{C})^{\perp}|=2^{m((k+1)n-((k+1)l_{0}+kl_{1}+\cdots+l_{k}))}.$
Since $\Phi$ is a bijection, then
$|\Phi(\mathcal{C}^{\perp})|=|\mathcal{C}^{\perp}|=2^{m((k+1)n-((k+1)l_{0}+kl_{1}+\cdots+l_{k}))}.$
 Hence $|\Phi(\mathcal{C})^{\perp}|=|\Phi(\mathcal{C}^{\perp})|.$  $\qquad\blacksquare$\\

\dse{4~~Code Construction \Rmnum{1 } } In this section, we construct
a new family of $2^{m}$-ary quantum codes by using
 dual-containing cyclic codes of odd length over $R.$  A fundamental link
between linear codes and quantum codes is given by the
Calderbank-Shor-Steane (CSS) construction. We first  recall some
definitions and notations.\\

 Let $f(x)=a_{k}x^{k}+a_{k-1}x^{k-1}+\cdots+a_{0}$ be a
polynomial in $R[x].$ Define the reciprocal polynomial of $f(x)$ as
$f^{*}(x)=x^{k}f(x^{-1})$, i.e.,
$f^{*}(x)=a_{0}x^{k}+a_{1}x^{k-1}+\cdots+a_{k}.$ Obviously,
$(f^{*}(x))^{*}=f(x)$ and  $(f(x)g(x))^{*}=f^{*}(x)g^{*}(x)$. If
there exists a unit $\varepsilon$ in $R$ such that $f(x)=\varepsilon
f^{*}(x)$, then $f(x)$ is called self-reciprocal over $R;$
otherwise, $f(x)$ is called non self-reciprocal over $R.$ If  $f(x)$
is an irreducible and non self-reciprocal divisor of $x^{n}-1,$ then
$f^{*}(x)$ is also an irreducible and non self-reciprocal divisor of
$x^{n}-1.$  Such $f(x)$ and $f^{*}(x)$ are called irreducible
reciprocal polynomial pairs. Let $f(x)$ be a monic factor of
$x^{n}-1,$ then we denote $\widehat{f}(x)=\frac{x^{n}-1}{f(x)}.$
Every cyclic code $\mathcal{C}$ has generator polynomial
in the following form.\\

\noindent\textbf{Theorem 4.1$^{[23]}$} \emph{Let $\mathcal{C}$ be a
cyclic code over $R$ of length $n$. Then there exists a unique
family of monic pairwise coprime polynomials $f_{0}(x), f_{1}(x),
f_{2}(x),\ldots,f_{k+1}(x)$ in $R[x]$ such that $f_{0}(x)f_{1}(x)
f_{2}(x)\cdots f_{k+1}(x)=x^{n}-1$ and $$\mathcal{C}=\left\langle
\widehat{f}_{1}(x),u\widehat{f}_{2}(x),u^{2}\widehat{f}_{3}(x),\ldots,
u^{k}\widehat{f}_{k+1}(x)\right\rangle$$ with
$|\mathcal{C}|=2^{m(\sum_{i=0}^{k}(k+1-i)degf_{i+1}(x))}$. Moreover
$$\mathcal{C^{\bot}}=\left\langle
\widehat{f}_{0}^{*}(x),u\widehat{f}_{k+1}^{*}(x),u^{2}\widehat{f}_{k}^{*}(x),\ldots,
u^{k}\widehat{f}_{2}^{*}(x)\right\rangle$$ with
$|\mathcal{C^{\bot}}|=2^{m(\sum_{i=0}^{k+1}idegf_{i+1}(x))}$. }\\

Now we obtain a sufficient and necessary condition for the existence
of dual-containing cyclic codes over $R$ by using  generator
polynomials of cyclic codes over $R.$\\

 \noindent\textbf{Theorem
4.2} \emph{ Let $\mathcal{C}$ be a cyclic code over $R$ of length
$n$. Then there exists a unique family of monic pairwise coprime
polynomials $f_{0}(x), f_{1}(x), f_{2}(x),\ldots,f_{k+1}(x)$ in
$R[x]$ such that $f_{0}(x)f_{1}(x) f_{2}(x)\cdots
f_{k+1}(x)=x^{n}-1$ and
$$\mathcal{C}=\left\langle
\widehat{f}_{1}(x),u\widehat{f}_{2}(x),u^{2}\widehat{f}_{3}(x),\ldots,
u^{k}\widehat{f}_{k+1}(x)\right\rangle.$$  Let $r_{i}(x)$ be the
product of irreducible and non self-reciprocal  divisors in
$f_{i}(x)$ which do not occur in pairs, for $i=2,3,\ldots,k+1.$
 Then  $\mathcal{C}^\bot \subseteq \mathcal{C}$ if and only
 if $(f_{0}(x)r_{2}(x)r_{3}(x)\cdots r_{k+1}(x))|f_{1}^{*}(x).$
 }\\

\noindent {\it \textbf{Proof}}~~ By Theorem 4.1, we obtain
$$
\mathcal{C^{\bot}}=\left\langle
\widehat{f}_{0}^{*}(x),u\widehat{f}_{k+1}^{*}(x),u^{2}\widehat{f}_{k}^{*}(x),\ldots,
u^{k}\widehat{f}_{2}^{*}(x)\right\rangle.$$ If $\mathcal{C}^\bot
\subseteq \mathcal{C},$ then there exists
$a(x)\in\mathbb{F}_{2^{m}}[x]$ such that
$\widehat{f}_{0}^{*}(x)=\widehat{f}_{1}(x)a(x).$ Let
$f_{i}(x)=b_{i}(x)r_{i}(x),$   where
$b_{i}(x)=\varepsilon_{i}b_{i}^{\ast}(x)$ with $\varepsilon_{i}\in
\mathbb{F}_{2^{m}}^{\ast}$ for $i=2,3,\ldots,k+1.$  Then
$$f_{1}^{*}(x)b_{2}^{*}(x)r_{2}^{*}(x)\cdots
b_{k+1}^{*}(x)r_{k+1}^{*}(x)= f_{0}(x)b_{2}(x)r_{2}(x)\cdots
b_{k+1}(x)r_{k+1}(x)a(x).$$  For $i=2,3,\ldots,k+1,$ each $r_{i}(x)$
is not a self-reciprocal polynomial, then $r_{i}(x)$ and
$b_{i}^{*}(x)r_{i}^{*}(x)$ are relatively coprime. Therefore,
$r_{i}(x)|f_{1}^{*}(x),$ for $i=2,3,\ldots,k+1.$  Since distinct
 $r_{i}(x)$ are relatively coprime, for $i=2,3,\ldots,k+1,$
it follows that $(r_{2}(x)r_{3}(x)\cdots r_{k+1}(x))|f_{1}^{*}(x).$
We know that $ f_{1}^{*}(x)f_{2}^{*}(x)\cdots
f_{k+1}^{*}(x)f_{1}(x)
 =f_{0}(x)f_{1}(x)f_{2}(x) \cdots  f_{k+1}(x)a(x)=
-f_{0}^{*}(x)f_{1}^{*}(x)f_{2}^{*}(x)\cdots f_{k+1}^{*}(x)\\ a(x),$
 then  $f_{0}(x)|f_{1}^{*}(x).$ Since  $f_{0}(x)$ and each $r_{i}(x)$ are
relatively coprime, we have $(f_{0}(x)r_{2}(x)r_{3}(x) \\ \cdots
r_{k+1}(x))  |f_{1}^{*}(x).$

On the other hand, if $(f_{0}(x)r_{2}(x)r_{3}(x)\cdots
r_{k+1}(x))|f_{1}^{*}(x),$ there exists $m(x) \in
\mathbb{F}_{2^{m}}[x]$ such that
$f_{1}^{*}(x)=f_{0}(x)r_{2}(x)r_{3}(x)\cdots r_{k+1}(x)m(x).$ Since
$f_{1}(x)$ and
 $f_{k+1}(x)$ are relatively coprime, there exist $s(x)$ and $t(x)$ in
 $\mathbb{F}_{2^{m}}[x]$ such that $f_{1}(x)s(x)+f_{k+1}(x)t(x)=1.$ Therefore
\begin{eqnarray*}
 u\widehat{f}_{k+1}^{*}(x)&=& uf_{0}^{*}(x)f_{1}^{*}(x)f_{2}^{*}(x)\cdots f_{k}^{*}(x)(f_{1}(x)s(x)+f_{k+1}(x)t(x))\\
&=& uf_{0}^{*}(x)(f_{0}(x)r_{2}(x)r_{3}(x)\cdots r_{k+1}(x)m(x))(b_{2}(x)r_{2}(x)\cdots b_{k}(x)r_{k}(x))^{*}f_{1}(x)s(x)\\
&&+uf_{0}^{*}(x)(f_{0}(x)r_{2}(x)r_{3}(x)\cdots r_{k+1}(x)m(x))(b_{2}(x)r_{2}(x)\cdots b_{k}(x)r_{k}(x))^{*}f_{k+1}(x)t(x)\\
&=&u\varepsilon_{2}\cdots \varepsilon_{k}f_{0}(x)f_{1}(x)f_{2}(x)\cdots f_{k}(x)f_{0}^{*}(x)r_{2}^{*}(x)\cdots r_{k}^{*}(x)r_{k+1}(x)m(x)s(x)\\
&&+u\varepsilon_{2}\cdots
\varepsilon_{k}f_{0}(x)f_{2}(x)f_{3}(x)\cdots
f_{k+1}(x)f_{0}^{*}(x)r_{2}^{*}(x)\cdots
r_{k}^{*}(x)r_{k+1}(x)m(x)t(x)\in \mathcal{C}.
\end{eqnarray*}
Similarly, we can prove that $u^{2}\widehat{f}_{k}^{*}(x) \in
\mathcal{C},\ldots, u^{k}\widehat{f}_{2}^{*}(x) \in \mathcal{C}.$
Furthermore

\begin{eqnarray*}
\widehat{f}_{0}^{*}(x)
&=&(f_{0}(x)r_{2}(x)r_{3}(x)\cdots r_{k+1}(x)m(x))(b_{2}(x)r_{2}(x)\cdots b_{k+1}(x)r_{k+1}(x))^{*}\\
&=& \varepsilon_{2}\cdots \varepsilon_{k+1}f_{0}(x)f_{2}(x)\cdots
f_{k+1}(x)m(x)r_{2}^{*}(x)r_{3}^{*}(x)\cdots r_{k+1}^{*}(x)\in
\mathcal{C}.
\end{eqnarray*}
Hence $\mathcal{C}^\bot \subseteq \mathcal{C}.$ This
completes the proof. $\qquad\blacksquare$\\

An important application of dual-containing codes is the
constructions of quantum codes. We denote by $[[n,l,d]]_q$ a $q$-ary
quantum code for $n$ qubits having $q^l$ codewords and minimum
distance $d$.  It is well known that quantum codes with parameters
$[[n,l,d]]_q$ must satisfy the quantum Singleton bound: $l \leq
n-2d+2$ (see Refs. [14-16]).
 A quantum code achieving this bound is called a quantum maximum-distance-separable (MDS) code. Quantum MDS codes are optimal.  A fundamental link between linear codes and quantum
codes is given by the CSS construction.\\

\noindent\textbf{Theorem 4.3$^{[1]}$} \emph{ (CSS construction) \
Let $\mathcal{C}$ and $\mathcal{C}'$  be two linear codes over
$\mathbb{F}_q$ with parameters $[n, t_1,d_1]_q$ and $[n,t_2,d_2]_q$.
If    $\mathcal{C}^\bot \subseteq \mathcal{C}',$ then there exists a
$q$-ary quantum code with parameters $\left[\left[n,  t_1+t_2-n, \\
 \min\left\{d_1, d_2 \right\}\right]\right]_q$.  Especially,  if $\mathcal{C}^\bot \subseteq \mathcal{C},$  then there exists a
$q$-ary quantum code with parameters $\left[\left[n, 2t_1-n,
d_1 \right]\right]_q$.} \\

 Now, based on dual-containing cyclic codes over $R,$  the Gray map and the CSS construction, we construct a new family
 of $2^{m}$-ary quantum codes. From Theorems 4.2
and 4.3,  we  directly get the following results.\\

\noindent\textbf{Theorem 4.4} \emph{Let $$\mathcal{C}=\left\langle
\widehat{f}_{1}(x),u\widehat{f}_{2}(x),u^{2}\widehat{f}_{3}(x),\ldots,
u^{k}\widehat{f}_{k+1}(x)\right\rangle$$  be a cyclic code over $R$
of length $n$,  type $\{l_{0},l_{1},\ldots,l_{k}\}$ and
 the minimum Gray distance $d_{G},$ where   $f_{0}(x), f_{1}(x),  f_{2}(x),\ldots, f_{k+1}(x)$ are a unique family of monic pairwise
coprime polynomials in $R[x]$ and $f_{0}(x)f_{1}(x) f_{2}(x)\cdots
f_{k+1}(x)=x^{n}-1.$  If there exist $r_{i}(x)$ are the product of
irreducible and non self-reciprocal divisors in $f_{i}(x)$ which do
not occur in pairs, for $i=2,3,\ldots,k+1$ and
$(f_{0}(x)r_{2}(x)r_{3}(x)\cdots r_{k+1}(x))|f_{1}^{*}(x),$ then
$\mathcal{C^{\bot}}\subseteq \mathcal{C},$ and there exists a
$2^{m}$-ary quantum code with parameters $[[(k+1)n,
2((k+1)l_{0}+kl_{1}+\cdots +l_{k})-(k+1)n,
d_{G}]]_{2^{m}}$.}\\

Let us use the following examples to illuminate the method of the
above construction.\\
 \noindent\textbf{Example 4.5} Consider a cyclic
code with length $15$ over
$\mathbb{F}_{2}+u\mathbb{F}_{2}+u^{2}\mathbb{F}_{2}+u^{3}\mathbb{F}_{2}.$
\\
Note that
$$x^{15}-1=(x-1)(x^{2}+x+1)(x^{4}+x+1)(x^{4}+x^{3}+1)(x^{4}+x^{3}+x^{2}+x+1)$$
over
$\mathbb{F}_{2}+u\mathbb{F}_{2}+u^{2}\mathbb{F}_{2}+u^{3}\mathbb{F}_{2}.$
Taking $ f_{1}(x)=1,
f_{2}(x)=(x^{2}+x+1)(x^{4}+x+1)(x^{4}+x^{3}+x^{2}+x+1),
f_{3}(x)=x^{4}+x^{3}+1, f_{4}(x)=1, f_{5}(x)=x-1. $
 Then $$\mathcal{C}=\left\langle
\widehat{f}_{1}(x),u\widehat{f}_{2}(x),u^{2}\widehat{f}_{3}(x),u^{3}\widehat{f}_{4}(x)\right\rangle$$
is a cyclic code over
$\mathbb{F}_{2}+u\mathbb{F}_{2}+u^{2}\mathbb{F}_{2}+u^{3}\mathbb{F}_{2}$
of length $15$,  type $\{10,4,0,1\},$
 the minimum Gray distance $4$ and  $\mathcal{C}^\bot \subseteq \mathcal{C}.$
 By Theorem 4.4, we can obtain a
 $Q=[[60,46,4]]_{2}$ quantum code, which meets the bound in Ref. 24. Thus, the
obtained quantum code is optimal.\\

\dse{5~~Code Construction \Rmnum{2 }}

{\upshape In this section, we will give an  another method to
construct quantum codes by using dual-containing cyclic codes of odd
length  over $R$. We recall some
definitions on the finite field $\mathbb{F}_{2^{m}}$ firstly. }\\
\noindent\textbf{Definition 5.1$^{[25]}$.}  \emph{Let
$B=\{\alpha_{1},\alpha_{2},\cdots,\alpha_{m}\}$ be a basis for
$\mathbb{F}_{2^{m}}$ over $\mathbb{F}_{2}.$   We call that $B$ is a
Trace Orthogonal Basis (TOB), if we have
\begin{eqnarray*}
\emph{Tr}(\alpha_{i} \alpha_{j}) = \left\{ {{\begin{array}{ll}
 {0}, & {\textrm{if}\mbox{ } \alpha_{i}\neq \alpha_{j}},\\
 {1}, & {\textrm{if}\mbox{ } \alpha_{i}= \alpha_{j}},\\
\end{array} }} \right .
\end{eqnarray*}
where $\emph{Tr}$ is the usual Trace function from
$\mathbb{F}_{2^{m}}$ to $\mathbb{F}_{2}.$}

 In this work, we let
$B=\{\alpha_{1},\alpha_{2},\cdots,\alpha_{m}\}$ be a self-dual basis
for $\mathbb{F}_{2^{m}}$ over $\mathbb{F}_{2}.$ Let $\mathcal{C'}$
be a cyclic code over $\mathbb{F}_{2^{m}}$ of length $N.$ For
$\textbf{c}'=(c'_{1},c'_{2},\ldots,c'_{N}).$ Define a map $\varphi$
by
$$\varphi : \mathbb{F}_{2^{m}}^{N}  \rightarrow  \mathbb{F}_{2}^{mN},$$
$$(c'_{1},c'_{2},\ldots,c'_{N}) \mapsto  (c_{11},c_{21},\ldots,c_{N1},
c_{12},c_{22},\ldots,c_{N2},\ldots,c_{1m},c_{2m},\ldots,c_{Nm}),$$
where $c'_{i}=\sum_{j=1}^{m}c_{ij}\alpha_{j}$ for $i=1,2,\ldots,N$
and $c_{ij}\in\mathbb{F}_{2}.$ Let $\textbf{a}\in
\mathbb{F}_{2}^{mN},$ with
$$\textbf{a}=(a_{11},a_{21},\ldots,a_{N1},
a_{12},a_{22},\ldots,a_{N2},\ldots,a_{1m},a_{2m},\ldots,a_{Nm})=(a^{(1)}|\cdots|a^{(m)}),$$
where $a^{(i)}\in \mathbb{F}_{2}^{N},$ for $i=1,2,\ldots,m.$ Let
$\sigma^{\otimes (m)}$ be the map from $\mathbb{F}_{2}^{mN}$ to
$\mathbb{F}_{2}^{mN}$ given by
$$\sigma^{\otimes(m)}(\textbf{a})=
(\tau(a^{(1)})|\cdots| \tau(a^{(m)})),$$ where $\tau$ is the usual
cyclic shifts defined in Section 2 on $\mathbb{F}_{2}.$  A code $C$
of length $mN$ over $\mathbb{F}_{2},$ if $\sigma^{\otimes (m)}(C)=C,
$ then the code $C$ is said to be a quasi-cyclic code of index
$m$ over $\mathbb{F}_{2}.$\\

 Now we need the following theorems for
 constructing a new family binary quantum codes from dual-containing cyclic
 codes of odd length
over
 $R.$\\

\noindent\textbf{Theorem 5.2$^{[13]}$} \emph{ Let $\mathcal{C}$ be a
cyclic codes over $\mathbb{F}_{2^{m}}$ with parameters $[N,
t,d]_{2^{m}}$. Then $\varphi(\mathcal{C})$ is a quasi-cyclic code of
index $m$ with parameters
$[mN, mt,d'\geq d]_{2}.$} \\

\noindent\textbf{Theorem 5.3} \emph{Let $\mathcal{C}$ be a cyclic
code over $R$ of length $n$,  type $\{l_{0},l_{1},\ldots,l_{k}\},$
 the minimum Gray distance $d_{G}$ and $\mathcal{C}^\bot \subseteq \mathcal{C}.$ Then $\varphi\circ\Phi(\mathcal{C})$ is
 a quasi-cyclic code of index
$m$ with parameters $[(k+1)mn,
m((k+1)l_{0}+kl_{1}+\cdots+l_{k}),d'\geq d_{G}]_{2}$ and
$\varphi\circ\Phi(\mathcal{C^{\bot}})\subseteq
\varphi\circ\Phi(\mathcal{C}),$
where $\circ$ means the composition of maps. }\\

\noindent {\it \textbf{Proof}}~~ Since $\mathcal{C}$ is a cyclic
code over $R$ of length $n$, type $\{l_{0},l_{1},\ldots,l_{k}\},$
and
 the minimum Gray distance $d_{G},$ $\Phi(\mathcal{C})$ is a cyclic
 code $\mathbb{F}_{2^{m}}$ with  parameters $[(k+1)n,
(k+1)l_{0}+kl_{1}+\cdots+l_{k},d_{G}]_{2^{m}}$. By Theorem 5.2, we
obtain that $\varphi \circ\Phi(\mathcal{C})$ is a quasi-cyclic code
of index $m$ with parameters $[(k+1)mn,
m((k+1)l_{0}+kl_{1}+\cdots+l_{k}),d'\geq d_{G}]_{2}.$ Now, we show
that $\varphi\circ\Phi(\mathcal{C^{\bot}})\subseteq
\varphi\circ\Phi(\mathcal{C}).$  Let
$c_{1}=\beta_{0}(c_{1})+u\beta_{1}(c_{1})+\cdots+u^{k}\beta_{k}(c_{1})$
and
$c_{2}=\beta_{0}(c_{2})+u\beta_{1}(c_{2})+\cdots+u^{k}\beta_{k}(c_{2})$
be any two elements in $\mathcal{C}$,
 where
 $\beta_{i}(c_{1})=a_{i1}\alpha_{1}+a_{i2}\alpha_{2}+\cdots+a_{im}\alpha_{m}$
and
$\beta_{i}(c_{2})=b_{i1}\alpha_{1}+b_{i2}\alpha_{2}+\cdots+b_{im}\alpha_{m},$
 with
 $a_{ij},b_{ij}\in \mathbb{F}_{2}$ for $i, \ j=0,1,\ldots,k.$ Since $\mathcal{C^{\bot}}\subseteq
\mathcal{C}$ and $\Phi$ is a bijection, it follows that
$\Phi(\mathcal{C^{\bot}})\subseteq \Phi(\mathcal{C}).$ Therefore,
$\sum_{i=0}^{t}\beta_{i}(c_{1})\beta_{t-i}(c_{2})=0,$ for
$t=0,1,\ldots,k;$ that is,
$\sum_{i,\mu,\lambda}a_{i,\mu}b_{t-i,\lambda}\alpha_{\mu}\alpha_{\lambda}=0,$
for $t=0,1,\ldots,k$ and $\mu, \lambda=1,2,\ldots,m.$
 Taking the trace
over $\mathbb{F}_{2},$   we get
$$\sum_{i,\mu,\lambda}a_{i,\mu}b_{t-i,\lambda}\textrm{Tr}(\alpha_{\mu}\alpha_{\lambda})=0.$$
Since $B=\{\alpha_{1},\alpha_{2},\cdots,\alpha_{m}\}$ is a self-dual
basis for $\mathbb{F}_{2^{m}}$ over $\mathbb{F}_{2}$, we get that
$\textrm{Tr}(\alpha_{\lambda}^{2})=1$ for  $\mu=\lambda$ and
$\textrm{Tr}(\alpha_{\mu}\alpha_{\lambda})=0$ for $\mu\neq \lambda.$
Therefore,
 $\sum_{i,\lambda}a_{i,\lambda}b_{t-i,\lambda}=0.$
This completes the proof. $\qquad\blacksquare$\\

Now, based on dual-containing cyclic codes over $R,$ the Gray map
and the trace map, we construct a new family
 of binary  quantum codes by using the CSS construction. From Theorems 4.2
and 5.3,  we  directly get the following results.\\

\noindent\textbf{Theorem 5.4} \emph{Let $$\mathcal{C}=\left\langle
\widehat{f}_{1}(x),u\widehat{f}_{2}(x),u^{2}\widehat{f}_{3}(x),\ldots,
u^{k}\widehat{f}_{k+1}(x)\right\rangle$$  be a cyclic code over $R$
of length $n$,  type $\{l_{0},l_{1},\ldots,l_{k}\}$ and
 the minimum Gray distance $d_{G},$ where   $f_{0}(x), f_{1}(x),  f_{2}(x),\ldots, f_{k+1}(x)$ are a unique family of monic pairwise
coprime polynomials in $R[x]$ and $f_{0}(x)f_{1}(x) f_{2}(x)\cdots
f_{k+1}(x)=x^{n}-1.$  Let $r_{i}(x)$ be the product of irreducible
and non self-reciprocal divisors in $f_{i}(x)$ which do not occur in
pairs, for $i=2,3,\ldots,k+1$ and $(f_{0}(x)r_{2}(x)r_{3}(x)\cdots
r_{k+1}(x))|f_{1}^{*}(x).$ Then $\mathcal{C}^\bot \subseteq
\mathcal{C}.$ Hence, there exists a binary quantum code with
parameters $[[(k+1)mn, 2m((k+1)l_{0}+kl_{1}+\cdots +l_{k})-(k+1)mn,
\geq d_{G}]]_{2}$.}\\

In the following, we give an example for construction a $4$-ary
quantum code and a binary quantum code by taking advantage of a
dual-containing cyclic code over
$\mathbb{F}_{2^{2}}+u\mathbb{F}_{2^{2}}.$\\

\noindent\textbf{Example 5.5}  \ Let
$\mathbb{F}_{2^{2}}=\{0,1,\omega,\omega^{2}\}$ be a finite field
with four elements, where $\omega^{2}=\omega+1.$ Consider a cyclic
code with length $21$ over $\mathbb{F}_{2^{2}}+u\mathbb{F}_{2^{2}}.$
\\
Note that\\
$x^{21}-1=(x-1)(x+\omega)(x+\omega^{2})(x^{3}+x+1)(x^{3}+x^{2}+1)(x^{3}+\omega
x+1)(x^{3}+\omega x^{2}+1)(x^{3}+\omega^{2} x+1)(x^{3}+\omega^{2}
x^{2}+1)$ \\ over $\mathbb{F}_{2^{2}}+u\mathbb{F}_{2^{2}}.$ Let $
f(x)=x^{3}+\omega^{2} x+1,
g(x)=(x-1)(x+\omega)(x+\omega^{2})(x^{3}+x^{2}+1)(x^{3}+\omega
x+1)(x^{3}+x+1)(x^{3}+\omega x^{2}+1)(x^{3}+\omega^{2} x^{2}+1),
h(x)=(x-1). $
 Then  $$\mathcal{C}=\left\langle
f(x)h(x),uf(x)g(x)\right\rangle$$ is a cyclic code over
$\mathbb{F}_{2^{2}}+u\mathbb{F}_{2^{2}}$ of length $21$,  type
$\{20,0,1\},$
 the minimum Gray distance $2$ and  $ \mathcal{C} \subseteq \mathcal{C}^{\bot}.$
 By Theorem 4.4, we can obtain a
 $Q'=[[42,40, 2]]_{4}$ quantum code, which is a quantum MDS code.  Thus, the obtained
quaternary quantum code is optimal.    On the other hand, we know
that $B=\{1,\omega\}$ is a self-dual basis for $\mathbb{F}_{2^{2}}$
over $\mathbb{F}_{2}.$ By Theorem 5.4, we can obtain a
 $Q''=[[84,80,d''\geq 2]]_{2}$ quantum code, which meets the bound in Ref.[24]. Thus,  the
obtained binary quantum code is good.

We list some quantum codes which can be constructed starting from
dual-containing cyclic codes over
$\mathbb{F}_{2^{2}}+u\mathbb{F}_{2^{2}}$ in Table \Rmnum{1 }.
Compared the parameters of quantum codes available in Refs. 12, 24,
26-28, we find that most of our obtained  quantum codes  have good
parameters.

\begin{table}[htbp]
{\small
\begin{center}
{\small{\bf TABLE I}~~ Quantum codes from dual-containing cyclic
codes over $\mathbb{F}_{2^{2}}+u\mathbb{F}_{2^{2}}$}
\begin{tabular}{c c c c c c c c c c c c c c c c c c c}
\hline
& Length & &  Minimum Gray distance & &  Type & & Quantum code \Rmnum{1 }& & Quantum code \Rmnum{2 }  \\
\hline
 & 7&  &$2$ &  & $\{6,0,1\}$ &  &  $[[14,12, 2]]_{4}$& &  $[[28,24,d'\geq 2]]_{2}$\\
  & 17& &$4$&  & $\{13,0,4\}$  &  &  $[[34,26, 4]]_{4}$& &  $[[68,52,d'\geq 4]]_{2}$\\
  & 31& &$4$&   & $\{25,0,6\}$  &  &  $[[62,50, 4]]_{4}$ & & $[[124,100,d'\geq 4]]_{2}$\\
 & 35& &$4$&   & $\{30,0,5\}$  &  &  $[[70,60, 4]]_{4}$& &  $[[140,120,d'\geq 4]]_{2}$\\
 & 43& &$5$&   & $\{36,0,0\}$  &  &  $[[86,78, 5]]_{4}$& &  $[[172,156,d'\geq 5]]_{2}$\\
\hline
\end{tabular}
\end{center}}
\end{table}

\noindent\textbf{Remark 5.5}  \ \ Compared with previously known
quantum codes in Refs. 12, 24,  26-28, some of our obtained quantum
codes are  optimal and new in Table \Rmnum{1 }. For quaternary case,
our obtained quantum codes  $[[14,12, 2]]_{4}$ and $[[86,78,
5]]_{4}$ are quantum MDS codes. Thus, they are optimal. Comparing
with the known quantum  code  $[[34,24, 4]]_{4}$  given in Ref.26,
our obtained quantum code $[[34,26, 4]]_{4}$ is optimal and new.
Comparing with the highest achievable minimum distance quantum
 codes $[[n, n-12, 4]]_{4}$  given in Ref.27(Corollary 3.2),  the obtained quantum code $[[62,50, 4]]_{4}$  has the same parameters and the obtained quantum code $[[70,60, 4]]_{4}$ is optimal and new.   For binary
case, we obtain the first three quantum codes, which nearly meet the
bound in
 Ref.24. Comparing
with the known quantum  code  $[[143,121, 3]]_{2}$ given in Ref.28,
our obtained quantum code  $[[140,120,d'\geq 4]]_{2}$ is good.
Furthermore, we obtain the  quantum code
 $[[172,156,d'\geq 5]]_{2}.$  Comparing with the known
quantum  codes  $[[170,130, 5]]_{2}$ and $[[178,154, 3]]_{2}$  given
in Ref.12, our obtained quantum code is good.
  The example shows that optimal and new quantum codes also can be
constructed from cyclic codes over finite rings.\\

\dse{6~~Conclusion}  Constructing quantum codes with good parameters
has been an important topic in quantum information. Many good
quantum codes were constructed from Hamming codes, BCH codes,
Reed-Solomon codes, Reed-Muller codes and algebraic geometry codes.
In this paper, two new families of  quantum codes have been
constructed by taking advantage of dual-containing cyclic codes over
the finite ring
$\mathbb{F}_{2^{m}}+u\mathbb{F}_{2^{m}}+\cdots+u^{k}\mathbb{F}_{2^{m}}.$
 The constructed quantum codes have good parameters. Furthermore, some of the constructed quantum codes are new.  The
results show that cyclic codes over finite rings are also a good
resource of constructing quantum codes. Our constructions are almost
theoretical, while it is
 very significant to discuss the implementation of these
 constructions in hardware. It would be very interesting to find a
 practical physical implementation for a quantum code from finite
 rings.

\rfne

\end{document}